\documentclass[twocolumn]{revtex4-1}
\usepackage{epsfig}
\usepackage{graphics}
\usepackage{amssymb}
\usepackage{textcomp}

\def\eq#1{(\ref{#1})}

\begin{document}

\title{On the holographic basis of Quantum Cosmology}

\author{C.A.S.Silva} 
\email{carlos.souza@ifpb.edu.br}
\affiliation{Instituto Federal de Educa\c{c}\~{a}o Ci\^{e}ncia e Tecnologia da Para\'{i}ba (IFPB),\\ Campus Campina Grande - Rua Tranquilino Coelho Lemos, 
671, Jardim Dinam\'{e}rica
I.}

\date{\today}

\begin{abstract}
Based upon the holographic principle, Jacobson demonstrated that 
the spacetime can be viewed as
a gas of atoms with a related entropy given by the Bekenstein-Hawking formula \cite{Jacobson:1995ab}.
Following this argument,
Friedmann equations can be derived by using Clausius relation to the 
apparent horizon of Friedmann-Robertson-Walker (FRW) universe 
\cite{Cai:2005ra}.  
Loop Quantum Gravity is a propose to description of the spacetime behavior in situations where its atomic characteristic arises. Among these situations,
the behavior of our universe near the Big Bang singularity is described by Loop Quantum Cosmology (LQC).
However, a derivation of the LQC equations based on the 
Bekenstein bound is lacking.
In this work, 
we obtain the quantum corrected Friedmann equations from the entropy-area relation
given by loop quantum black holes (LQBH), setting a still absent connection between holographic and 
LQC descriptions of the cosmos. Connections with braneworld cosmology have been also addressed.
\end{abstract}


\maketitle

%
%

Among the results coming from black hole thermodynamics, we have the Bekenstein-Hawking formula, where the 
entropy of a black hole is given as proportional to its horizon area: $S = A/4\hslash G$ \cite{Bekenstein:1973ur}.
Behind the simplicity of this expression, lies a deep intersection between two theories that remain at odds until now,
gravity and quantum mechanics. Interestingly,
Bekenstein-Hawking formula is one of the few places in physics
where the Newton's gravitational constant $G$ meets the Planck constant $\hslash$. 
In fact, String theory and
Loop Quantum Gravity (LQG) have shown that the origin of
the black-hole thermodynamics must reside in the
quantum structure of the spacetime \cite{Rovelli:1996dv, Ashtekar:1997yu, Strominger:1996sh}. 
Moreover, Bekenstein-Hawking formula
consists in the basis
of the holographic principle which sets how many degrees of freedom there are in nature at
the most fundamental level 
\cite{'tHooft:1993gx, Susskind:1994vu, Bousso:2002ju}, and has been claimed  
as a guide principle to quantum gravity. Holographic principle has led also to the development of the AdS/CFT correspondence, which applications go beyond the conventional
use in string theory \cite{Aharony:1999ti} and M-theory \cite{Aharony:2008ug}, but includes its aptness for understanding condensed matter systems
\cite{Jensen:2010em, Donos:2010ax}.

Hooked up with the results above, we have another signal of the relationship between black hole 
thermodynamics and the quantum structure of spacetime. Assuming the proportionality between entropy and horizon area, 
Jacobson derived the Einstein field equations
by using the fundamental Clausius relation, $\delta Q = TdS$, connecting heat, temperature and entropy \cite{Jacobson:1995ab}. 
The idea behind this result is to demand that the Clausius relation holds for all the local Rindler causal horizon through each spacetime point, with
$\delta Q$ and $T$ interpreted, respectively, as the energy flux and Unruh temperature seen by an accelerated observer just inside 
the horizon. The most important lesson which brings from this result is that the spacetime
can be viewed as a gas of atoms with a related entropy given by the Bekenstein-Hawking formula, and the 
Einstein's field equation is nothing, but an equation of state of this gas.
Such an interpretation of spacetime was later reinforced by Padmanabhan who linked the
macroscopic description of spacetime, described by Einstein equations,
to microscopic degrees of freedom by assuming the 
principle of
equipartition of energy \cite{Padmanabhan:2010xh}.


Jacobson's and Padmanabhan's results are among the main arguments in favor of the holographic hypothesis, and has given rise
to several works which have strengthened the thermodynamical interpretation
of Einstein's equations .
Actually, it has been shown
that the susceptibility of gravitational fields to a thermodynamic behavior occurs not only in Einstein's gravity
but also in a wide assortment of theories. (For a review and a broad list of references see \cite{Padmanabhan:2009vy}).
Among this, one interesting result is that Friedmann equations can be derived by the use of the Clausius relation to the 
apparent horizon of FRW universe, in which entropy is assumed to be proportional to its horizon area. 
This works not only in Einstein gravitational theory, but also in 
Gauss-Bonnet and Lovelock gravities \cite{Cai:2005ra}.
All these facts confirms that the connection between gravity and thermodynamics demonstrated
by Jacobson must be relevant to the understanding of the nature of gravity.
Particularly, the derivation of Friedmann equations from the thermodynamical Jacobson's argument
give support to a new perspective, introduced by Fischller, Suskind \cite{Fischler:1998st}
and Banks \cite{Banks:2001px}, where holography has been built as a guide principle to cosmology.

As the holographic hypothesis has established a deep connection between classical general relativity and
thermodynamics, one could think about the possibility of a connection between 
quantum gravity and a statistical theory of spacetime. However, 
even though holography constitutes an important argument in support of the idea that spacetime 
has an atomic structure, in the same way that Boltzmann kinetic theory of gases foresee the 
reality of atoms of matter,
it does not tell us how such a structure would be. In this way, holography asks for 
a theory that provides a microscopical
description of spacetime. LQG, on the other hand, proposes a way 
to model the behavior of spacetime in situations where its atomic characteristic arises. Among these situations,
the behavior of our universe near the Big Bang singularity is described by Loop Quantum Cosmology (LQC), which is
a model for the universe based on the quantization methods of 
LQG \cite{Bojowald:2006da}. 
This description of cosmology which takes into account effects of quantum gravity 
has become very popular during the last decade, because it allows making contact to observation 
\cite{Barrau:2013ula, Ashtekar:2015dja}. Moreover, a possible duality between LQC and braneworld cosmology
has been setting,
even though these 
two descriptions of the universe remains at odds in relation of the right signal of 
the quantum corrections in the friedmann equations \cite{Singh:2006sg}.

The best known finding in LQC is the resolution of the Big Bang singularity, since there are long
standing prospects that the General Relativity initial singularity
must be solved in the context of a quantum gravity theory. In the case of LQC, the Big Bang singularity is naturally replaced by a bounce that takes place when the curvature becomes strong 
(reaches the Planckian regime). In this point, the universe density does not 
becomes infinite anymore, but assumes a maximum finite critical value. 
This in return has consequences for example for the spectrum of 
primordial gravitational waves. Moreover, as has been shown by 
\cite{Singh:2009mz}, the existence of an upper bound on the energy density of the mater in LQC, which translates into an upper bound for the trace of extrinsic curvature, not only results
in the resolution of Big Bang singularity, but also
in a generic resolution
of all strong curvature singularities and finiteness of spacetime curvature for flat isotropic LQC.

All these facts have put LQC in a current exciting status. However, in order to have a statistical description
of the universe evolution based on LQC, its is necessary to answer  
a question that remains open until now:
what is the correct way to count the spacetime states in agreement with the LQC
description of the universe? 
To find out the correct answer to this question, one must to remember that it
must lie in the holographic principle in scenarios where gravity has a fundamental rule,
as in the context of LQC \cite{Bousso:2002ju}. 
In this way,  a holographic prescription to LQC becomes
necessary.

The issue of the relation between LQC and holography has been investigated by Cai et al \cite{Cai:2008ys},
by the use of Jacobson formalism, where a logarithmic quantum corrected Bekenstein-Hawking formula 
which arises in the context of LQG \cite{Meissner:2004ju} has been taken into account.
However, this attempt to obtain the LQC equations from
Bekenstein-Hawking entropy, has led to 
entropic corrected Friedmann equations which give a
different scenario from LQC ones. Actually, the worse conclusion from
the analysis done by Cai et al is that a bounce does not 
occurs and the Big-Bang singularity is not resolved, which establishes a breakup between the 
description of the spacetime behavior near the Big Bang and 
the way how its degrees of freedom are counted in the context of 
LQG. By the way, in the work by Cai et al, a relation between black hole entropy and horizon area is found out in agreement
with LQC. However, this relation have no physical justification, but consist just in a mathematical
result found from their own LQC equations.
In front of this difficult, 
the important issue of the description of
the thermodynamical evolution of the universe through its different eras, may be being 
built on sand in the context of LQC.

%

In this work, we shall follow an alternative way to get quantum corrections to Bekenstein-Hawking formula 
that arises 
in the context of loop black holes \cite{Modesto:2008im, 
Carr:2011pr, s.hossenfelder-grqc12020412}, to obtain 
obtain the correct count of spacetime degrees of freedom in order to reproduce the LQC description of the cosmos.
It will pave the way to a 
statistic mechanical prescription to LQC based on the
holographic principle. 
In this paper, we shall use $\hslash = c = k_{B} = G = 1$.

\emph{Loop quantum black holes \textthreequartersemdash}
A loop black hole, also called self-dual black hole, consists
in a quantum gravity corrected Schwarzschild black hole that appears from a simplified model of LQG
by the use of semiclassical
tools in the minisuperspace quantization approach.
The loop quantum black hole scenario is described  by a quantum gravitationally corrected Schwarzschild metric, 
given by $ds^{2} = - G(r)dt^{2} + F^{-1}(r)dr^{2} + H(r)(d\theta^{2} + \sin^{2}\theta d\phi^{2})$,
where the metric functions are given by

\begin{eqnarray}
G(r) &=& \frac{(r-r_{+})(r-r_{-})(r+r_{*})^2}{r^{4}+a_{0}^{2}} \; , \nonumber \\
F(r) &=& \frac{(r-r_{+})(r-r_{-})r^{4}}{(r+r_{*})^{2}(r^{4}+a_{0}^{2})} \; \hspace{1mm};  \hspace{1mm} 
H(r) = r^{2} + \frac{a_{0}^{2}}{r^{2}} \; . \label{h-form}
\end{eqnarray}

\noindent In this scenario, we have the presence of two horizons - an event horizon localized at $r_{+} = 2m$
and a Cauchy horizon localized at $r_{-} = 2mP^{2}$. Furthermore, we have that
$r_{*} = \sqrt{r_{+}r_{-}} = 2mP$, where $P$ is the polymeric function given by 

\begin{equation}
P = \frac{\sqrt{1+\epsilon^{2}} - 1}{\sqrt{1+\epsilon^{2}} +1} \;\; ; \;\; a_{0} = 
\frac{A_{min}}{8\pi}\;,
\end{equation}

\noindent and $A_{min}$ is the minimal value of area in Loop Quantum  Gravity. 

In the metric above, r is only asymptotically the usual radial coordinate
since $g_{\theta\theta}$ is not just $r^{2}$. 
A more physical radial coordinate is obtained from the form of the function $H(r)$ in the metric \eq{h-form}

\begin{equation}
R = \sqrt{r^{2}+\frac{a_{0}^{2}}{r^{2}}} \label{phys-rad} \; ,
\end{equation}

\noindent in the sense that it measures the proper circumferential distance.
Moreover,
the parameter $m$ in the solution is
related to the ADM mass $M$ by $M = m(1 + P )^{2}$.


An interesting property of LQBH is the property of self-duality. This property says that if one introduces the new coordinates $\tilde{r} = a_{0}/r$ and $\tilde{t} = t r_{*}^{2}/a_{0}$, with $\tilde{r}_{\pm} = a_{0}/r_{\mp}$ the metric
preserves its form. The dual radius is given by $r_{dual} = \tilde{r} = \sqrt{a_{0}}$ and corresponds to the minimal possible surface element.
Moreover, since the Eq. \eq{phys-rad} can be written as $R = \sqrt{r^{2}+\tilde{r}^{2}}$, 
it is clear that the solution 
contains another asymptotically flat Schwazschild region rather than
a singularity in the limit $r\rightarrow 0$. This new region corresponds to a Planck-sized wormhole. 
%
%



The derivation of the black hole's thermodynamical properties from the LQBH metric 
proceeds in the usual way. The Bekenstein-Hawking temperature $T_{BH}$
can be obtained by the calculation of the surface gravity $\kappa$ by $T_{BH} = \kappa/2\pi$, with

\begin{equation}
\kappa^{2} = - g^{\mu\nu}g_{\rho\sigma}\nabla_{\mu}\chi^{\rho}\nabla_{\nu}\chi^{\sigma} = 
- \frac{1}{2}g^{\mu\nu}g_{\rho\sigma}\Gamma^{\rho}_{\mu 0} \Gamma^{\sigma}_{\nu 0}\;,
\end{equation}

\noindent where $\chi^{\mu} = (1,0,0,0)$ is a timelike Killing vector and $\Gamma^{\mu}_{\sigma\rho}$ are the connections coefficients.

By connecting with the metric, one obtains that the loop quantum black hole temperature is given by

\begin{equation}
T_{H} =  \frac{(2m)^{3}(1-P^{2})}{4\pi[(2m)^{4} +a_{0}^{2}]}\;. \label{lb-temperature}
\end{equation}

\noindent It is easy to see that one can recover the usual Hawking temperature in the limit of large masses. However, differently from
the Hawking case, the temperature \eq{lb-temperature}
goes to zero for $m \rightarrow 0$. In this point, we remind that the black hole’s ADM mass
$M = m(1 + P )^2 \approx m$, since $P \ll 1$.

The black hole's entropy can be found out by making use of the
thermodynamical relation $S_{BH} = \int dm/T (m)$.

\begin{equation}
S = \frac{4\pi (1+P)^{2}}{(1-P^{2})}\Big[\frac{16m^{4} - a_{0}^{2}}{16m^{2}}\Big]\;.
\end{equation}

Moreover, one can obtain an expression for the black hole entropy in terms of its area \cite{s.hossenfelder-grqc12020412}

\begin{equation}
S =  \pm \frac{\sqrt{A^{2} - A_{min}^{2}}}{4}\frac{(1+P)}{(1-P)} \label{entropy-area}
\end{equation}

\noindent where we have set the possible additional constant to zero. S is positive for $m > \sqrt{a_{0}}/2$
and negative otherwise.
The double possibility in the signal of the loop black hole entropy is related with
the two possible physical situations that arise from loop quantum black hole structure, where the event horizon can
be outside or inside of the wormhole throat 
\cite{Carr:2011pr}.

The thermodynamics properties of LQBH has been also obtained through the Hamilton-Jacobi
version of the tunneling formalism \cite{Silva:2012mt}. By the use of this formalism, back-reaction effects 
could be included. Moreover, extensions of the loop quantum black hole solution to scenarios where charge and 
angular momentum are preset can be found in \cite{Caravelli:2010ff}. 
The issue of information loss has been also addressed in the context of loop black holes.
In this case, it has been pointed that, in this framework,
the problem of information loss by black holes could be relieved \cite{Alesci:2011wn, Silva:2012mt, Alesci:2012zz}. This result may be related with the absence of
a singularity in the loop black hole interior, and consists in a forceful result in benefit of this approach.

\emph{Quantum corrected Friedmann equation from LQBH \textthreequartersemdash} 
In despite of its complexity, according to the cosmological principle, our universe 
can be considered, at very large scale, as
homogeneous and isotropic.
Based on this simplifying assumption, the Friedmann
equations are a set of equations that govern the expansion of the universe in the context of General Relativity.
They were first derived by Alexander Friedmann in $1922$ from Einstein's field equations of gravitation for 
the Friedmann-Lema\^{\i}tre-Robertson-Walker metric and a perfect fluid with a given mass density 
and pressure \cite{friedmann-zphys10}. 

The Friedmann equation of an uniform cosmology is typically written in the form

\begin{equation}
H^{2} + \frac{k}{R^{2}} = \frac{8\pi}{3}\rho\; . \label{classical-friedmann}
\end{equation}

\noindent In the equation above, $H$ is the Hubble parameter, $R$ is a scale factor of the universe, 
$\rho$ is the energy density, and $k$ is a dimensionless constant ̇
related to the curvature of the universe. The Hubble parameter is defined as $H = \dot{R}/R = \dot{a}/a$, ̇
where $a$ is the dimensionless scale factor of the universe given by $a = R/R_{0}$ and $R_{0}$ is the scale factor
of the universe at some canonical time $t_{0}$ . An example of $R_{0}$ is the average distance between
galaxies.

Friedmann equations must incorporate quantum corrections in order to explain the evolution 
of the universe in the stages close to the Big Bang singularity, where the spacetime must have a quantum 
behavior. In fact, quantum corrections to Friedmann equations have been found out in the context of Loop Quantum
Gravity, where the quantum corrections imply in an additional contribution in the density term.
Due to the additional term in the Friedmann equation, a quantum
bounce replaces the Big Bang singularity when the density of the universe assumes a critical value given by
$\rho_{crit} = \sqrt{3}/(32\pi G^{2} \gamma^{3})$, where $\gamma$ is 
the Barbero-Immirzi parameter \cite{Ashtekar:2008gn, Bojowald:2006da}.

In this paper, we shall start from the assumption that the entropy associated with the apparent horizon 
of the universe is related with its area by the modified entropy-area relation \eq{entropy-area}, 
to obtain quantum gravity corrections to Friedmann equations. To do this, we shall follow the
procedure developed in \cite{Cai:2005ra} and \cite{Cai:2008ys}.

The FRW universe is described by the following metric

\begin{eqnarray}
ds^{2} &=& - dt^{2} + a(t)^{2}\Big(\frac{dr^{2}}{1- kr^{2}} + r^{2}d\Omega^{2}_{2}\Big) \nonumber \\
        &=& h_{ab}dx^{a}dx^{b} + \tilde{r}^{2}d\Omega^{2}_{2}\label{frw-metric} \; ,
\end{eqnarray}

\noindent where $h_{ab} = diag (-1, a^{2}/(1- kr^{2}))$ and $\tilde{r} = a(t)r$.
Moreover, the radius of the apparent horizon is given by

\begin{equation}
\tilde{r}_{A} = \frac{1}{\sqrt{H^{2} + k/a^{2}}} \label{ap.hor-rad}\;. 
\end{equation}


Now, let us suppose that the energy-momentum tensor $T_{\mu\nu}$ of the matter in universe has the form of a perfect
fluid: 

\begin{equation}
T_{\mu\nu} = (\rho + p)U_{\mu}U_{\nu} + p g_{\mu\nu}\;.
\end{equation}

The energy conservation law leads to the continuity equation

\begin{equation}
\dot{\rho} + 3H(\rho + p) = 0\;. \label{continuity-eq}
\end{equation}

In this point, we shall define the work density $W$ and the energy-supply vector $\psi$ as

\begin{equation}
W = -\frac{1}{2}T^{ab}h_{ab}\; ;
\end{equation}
\noindent and

\begin{equation}
\psi_{a} = T^{b}_{a}\partial_{b}\tilde{r} + W\partial_{a}\tilde{r} \; .
\end{equation}

\noindent We shall have, in our case

\begin{equation}
W = \frac{1}{2}(\rho - p)\;;
\end{equation}
\noindent and

\begin{equation}
\psi_{a} =  - \frac{1}{2}(\rho + p)H\tilde{r}dt + \frac{1}{2}(\rho + p)adr \; .
\end{equation}

From the expressions above, we can compute the amount of energy going through 
the apparent horizon during the time interval
$dt$ as \cite{Cai:2005ra}

\begin{equation}
\delta Q = - A\psi = A(\rho+p)H\tilde{r}_{A}dt \; ,
\end{equation}

\noindent where $A = 4\pi\tilde{r}^{2}_{A}$.

As have been emphasized by Cai et al \cite{Cai:2008ys}, the horizon temperature is 
completely determined by the spacetime metric, 
independently of gravity theories. 
On the other hand, the horizon entropy depends on gravity theory we are considering.
The temperature associated with the apparent horizon is given by 
$T = (2k_{B}\pi\tilde{r}_{A})^{-1}$, which was obtained in the reference \cite{Cai:2008gw} through tunneling methods.
On the other hand, the apparent horizon entropy will be given by the Eq. \eq{entropy-area}. In other words,
only the entropy-area relation will be changed.

With all these results in our hands, using the Clausius relation
$\delta Q = TdS$, we can reach

\begin{equation}
\dot{H} - \frac{k}{a^{2}} =  \mp 4\pi \frac{(1-P)}{(1+P)}\frac{\sqrt{A^{2}-A_{min}^{2}}}{A}(\rho + p)\;. 
\label{friedmann1}
\end{equation}

\noindent In order to obtain the Friedmann equation above we have used the relation

\begin{equation}
\dot{\tilde{r}}_{A} = -H\tilde{r}^{3}_{A}\Big(\dot{H} - \frac{k}{a^{2}}\Big)\;.
\end{equation}

\noindent Now, using the continuity Eq. \eq{continuity-eq}, we can find

\begin{equation}
\frac{8\pi}{3}\frac{d\rho}{dt} = 
\pm \frac{(1+P)}{(1-P)}\frac{A}{\sqrt{A^{2} - A_{min}^{2}}} \frac{d(H^{2}+k/a^{2})}{dt} \; .
\end{equation}

\noindent Integrating the equation above yields

\begin{equation}
H^2 + \frac{k}{a^2} = \frac{4\pi}{A_{min}}\cos(\Lambda)\;, \label{friedmann2}
\end{equation}

\noindent where $\Lambda = \pm\Big[\frac{2A_{min}}{3}\frac{(1-P)}{(1+P)}\rho - \alpha\Big]$, and $\alpha$ is a phase 
constant. 

The Eqs. \eq{friedmann1} and \eq{friedmann2} above consist in the quantum version of the Friedmann equations 
which appear in the context of LQBH. As we can see, the quantum corrections present in these
equations, inherited from the loop quantum black hole entropy-area relation, implies in a quantum effective density term 
which is a harmonic function of the classical density. A very important consequence of this result is that  
the quantum corrected Friedmann equations bring us a scenario where the Big Bang initial 
singularity does not exist anymore,
but is replaced by a bounce at a point where the universe density gets a critical critical value, 
as occurs in LQC.

\emph{Relation with usual semiclassical LQC and braneworld cosmology \textthreequartersemdash }
In this point, we shall put all quantum corrections
that appears in the Eq. \eq{friedmann2} on the right-hand side. 
To do this, we shall write this equation in terms of the classical radial coordinate, i.e, 
in terms of the radial coordinate $\tilde{r}_{A}'$ in the relation $
\tilde{r}_{A} = \sqrt{\tilde{r'}_{A}^{2} + a_{0}^{2}/\tilde{r'}_{A}^2}$ (See the Eq. \eq{phys-rad}).
Moreover, we shall make the phase constant in the Eq. \eq{friedmann2} explicit, 
in order to obtain the correct value of the universe critical density, i.e, 
the universe density when the bounce occurs. 

In this way, using the Eq. \eq{ap.hor-rad}, we have that

\begin{eqnarray}
H^{2} + \frac{k}{a^{2}} &=& \frac{H'^{2} + \frac{k}{a'^{2}}}{1+a_{0}^{2}(H'^{2} + \frac{k}{a'^{2}})^2} \nonumber \\
&=&  \frac{4\pi}{A_{mim}}\cos(\Lambda) \; ,
\end{eqnarray}

\noindent where the commas denote that the functions depend on $r'$.

Solving the equation above, we find out two identical solutions given by

\begin{equation}
H'^{2} + \frac{k}{a'^{2}} =  \frac{1 + \sin\Big[(2/3)A_{min}\frac{(1-P)}{(1+P)}\rho - \alpha\Big]}
{a_{0}\cos\Big[(2/3)A_{min}\frac{(1-P)}{(1+P)}\rho - \alpha\Big]} \; .
\end{equation}

The expansion of the right-hand side of the equation above give us

\begin{equation}
H'^{2} + \frac{k}{a'^{2}} = A(\alpha)\rho^{2} + B(\alpha)\rho +C(\alpha) + 
\ensuremath{\mathcal{O}(\geq A_{min}^{2}}) \; ,
\end{equation}
\noindent where

\begin{eqnarray}
A(\alpha) = &-&\frac{32\pi}{9}A_{min}\frac{(1-P)^{2}}{(1+P)^{2}}\sec(\alpha)\Big(\tan^{2}(\alpha) \nonumber \\
&-&\tan(\alpha)\sec(\alpha)+ \frac{1}{2}\Big) \; , \nonumber \\
B(\alpha) = &&\frac{16\pi}{3}
\frac{(1-P)}{(1+P)}\sec(\alpha)(\sec(\alpha) - \tan(\alpha)) \; , \nonumber \\
C(\alpha) = &&\frac{8\pi}{A_{min}}(\sec(\alpha) - \tan(\alpha)) \; , \label{coefficients}
\end{eqnarray}


\noindent and the terms of order $\ensuremath{\mathcal{O}(\geq A_{min}^{2}})$ can be disconsidered front of the 
other ones.

In order to fix correctly the phase constant in the equation \eq{coefficients},
with shall take the simple redefinition of the universe density $\rho \rightarrow (\varrho)^{-1}(\beta\rho-\delta)$, 
with $\varrho = (B+2A)/A$, $\beta = B+2A$, and $\delta = (B+2A)/2A)(1\pm\sqrt{1+4AC/(B+2A)}$, one obtains:

%
%


 \begin{eqnarray}
H^{2} + \frac{k}{a^{2}} = \frac{8\pi G}{3}\rho\Big(1 - \frac{\rho}{\varrho(\alpha)}\Big) \label{lqc-eq}\; ,
\end{eqnarray}



\noindent where the signal of $\varrho(\alpha)$ depends of the choice of the phase constant. 

If we appropriately choice the phase constant as 
$\alpha \approx 0.7805\;\textrm{rad}$, we obtain, for
$\gamma = 0.2375$, $\varrho = \rho_{c} = \sqrt{3}/(32\pi\gamma^{3})$, the critical density value in LQC.
In this way, usual semiclassical LQC Friedmann equation can be recovered as an approximation of the equation of state
\eq{friedmann2}.

On the other hand, a very interesting fact occurs if one takes values of the phase constant in a way that
$\varrho(\alpha) < 0$. In this case, we can
rewrite it as $\varrho = -2\sigma$, for $\sigma > 0$, and then obtain

 \begin{eqnarray}
H^{2} + \frac{k}{a^{2}} = \frac{8\pi G}{3}\rho\Big(1 + \frac{\rho}{2\sigma}\Big) \label{lqc-eq}
\end{eqnarray}

\noindent which corresponds to the braneworld cosmology Friedmann equation with $\sigma$ as the 
brane tension \cite{Maartens:2003tw}.
As a matter of fact, the duality between LQC and
braneworld cosmology has been addressed in \cite{Singh:2006sg}.
From the results above, LQC and braneworld cosmology equations can emerge from the equation
\eq{friedmann2}, differing only by a phase constant.

%
%
%

%

%
%
%
%
%

\emph{Summary and outlook \textthreequartersemdash}
Quantum corrected Friedmann equations has been derived from the modified Bekenstein-Hawking
formula to black hole entropy solution that arises in the context of
LQBH. 
The compatibility with the results of standard LQC, following the Jacobson's approach
\cite{Jacobson:1995ab} might suggest that the black hole entropy counting
performed in loop black hole's scenario may be on a more solid footing than other counting procedures
such as those given in other approaches like \cite{Meissner:2004ju}. \\
\indent The usual semiclassical LQC, as well as braneworld cosmology, equations are recovery when one discards the
upper orders of quantum corrections through of an appropriate choice of a phase constant.
In despite of the fact that braneworld Friedmann equations do not admit a bounce, 
its occurrence is ensured by
\eq{friedmann2}. Since this equation is an equation of state for the cosmological evolution
of spacetime, semiclassical LQC, and more surprisingly braneworld cosmology,
would appear as a thermodynamical effect whose origin would lies in the atomic structure
of spacetime described by LQG. This result could be a sign of a deep connection between LQG and string theory
where the later could emerge from the first as a thermodynamic effect, consequence of
a quantum spacetime measurement mechanism.

\bibliography{basename of .bib file}


\end{document}